\documentclass[10pt,twocolumn,letterpaper]{article}

 \usepackage[pagenumbers]{iccv} % To force page numbers, e.g. for an arXiv version

\usepackage{verbatim}
\definecolor{iccvblue}{rgb}{0.21,0.49,0.74}
\usepackage[pagebackref,breaklinks,colorlinks,allcolors=iccvblue]{hyperref}

\def\paperID{*****} % *** Enter the Paper ID here
\def\confName{ICCV}
\def\confYear{2025}

\title{EchoNet-Quality: Denoising Echocardiograms via Deep Generative Modeling of Ultrasound Noise}

\author{
David Choi$^{1}$\hspace{0.5em}Milos Vukadinovic$^{1,2,3}$\hspace{0.5em}Bryan He$^{3,4}$\hspace{0.5em}Christina Binder$^{5}$\hspace{0.5em}Yuki Sahashi$^{1}$\hspace{0.5em}David Ouyang$^{1,3}$\\
\textsuperscript{\rm 1}Department of Cardiology, Cedars-Sinai Medical Center \\
\textsuperscript{\rm 2}Department of Bioengineering, University of California Los Angeles \\
\textsuperscript{\rm 3}Department of Cardiology, Kaiser Permanente \\
\textsuperscript{\rm 4}Department of Computer Science, Stanford University \\
\textsuperscript{\rm 5}Department of Internal Medicine, Division of Cardiology, Medical University of Vienna \\
\small{\texttt{\{david.choi2, yuki.sahashi\}@cshs.org, milosvuk@ucla.edu, christina.binder@meduniwien.ac.at,}}\\
\small{\texttt{bryanhe@cs.stanford.edu, david.ouyang@kp.org}}
}

\begin{document}
\maketitle
\begin{abstract}
Echocardiography (echo), or cardiac ultrasound, is the most widely used imaging modality for cardiac form and function due to its relatively low cost, rapid acquisition time, and non-invasive nature. However, ultrasound acquisitions are often limited by artifacts and noise that hinder diagnostic interpretation in clinical settings. Existing methodologies for denoising echos consist solely of traditional filtering-based algorithms or deep learning methods developed on radio-frequency (RF) signals which prevents clinical applicability and scalability. To address these limitations, we introduce the first deep generative model capable of simulating ultrasound noise developed on B-mode data. Using this generative model, we develop a synthetic dataset of paired clean and noisy echo images to train a downstream model for real-world image denoising and demonstrate state-of-the-art performance in both internal and external experiments. In both held-out test sets, our method results in echo images with higher gCNR in comparison to noisy image counterparts and images derived from a comparable method which is consistent with provided visual comparisons. Our experiments showcase the potential of our method for future clinical use to improve the quality of echo acquisitions. To encourage further research into the field, we release our source code and model weights at \texttt{https://github.com/echonet/image\char`_quality}.
\end{abstract}    
\section{Introduction}
\label{sec:intro}

\begin{figure*}[!ht]
 \centering
 \includegraphics[width=0.99\linewidth]{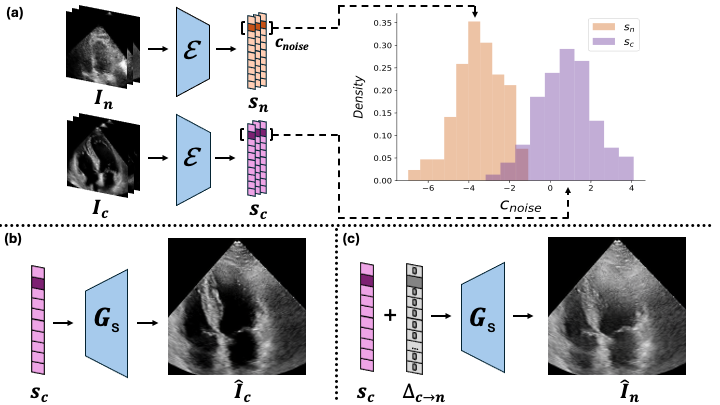}
 \caption{Overview of our noise simulation method. (a) Project physician-labeled noisy and clean images to latent embeddings and identify noise channel via statistical separation. (b) Find latent embedding corresponding to some clean image. (c) Manipulate noise channel in clean latent embedding to generate noisy image.}
 \label{figure:overview}
\end{figure*}

Transthoracic echocardiography (TTE) is a non-invasive imaging modality used for monitoring cardiac function and diagnosing abnormalities. In many cases, TTE is the first and most common option for screening heart diseases such as endocarditis, valvular heart disease, etc. Unfortunately, TTE acquisitions are prone to a form of noise referred to as acoustic clutter caused by reverberations in echogenic structures such as subcutaneous fat, bones, and lungs blocking the propagation of ultrasound waves \cite{reverb}. This can result in low visual distinction between cardiac chambers and tissue in B-mode echo which pose challenges for clinical diagnosis \cite{diagnosis}, measurement reliability \cite{reliability}, and post-formation image processing \cite{processing}. Altogether, these challenges motivate the development of techniques for denoising echo.

Recent advances in the field of generative modeling have allowed models to learn complex and high-dimensional distributions of data. This has led to data-driven methods achieving far greater capabilities for domain translation and inverse problems in the field of medical imaging by leveraging generative models e.g. Generative Adversarial Networks (GANs) \cite{gan} as data priors. Notably, developments in GAN architectures and techniques e.g.\ StyleGAN \cite{StyleGAN} have demonstrated the capability to induce semantic changes in images by learning a well-structured latent space representation of data, projecting data into latent embeddings, and interpolating in semantically relevant directions \cite{editing1, editing2, editing3}.

To solve the problem of denoising echo images, we adopt the StyleGAN architecture to learn a latent space representation of echo images and leverage it to encode the intrinsic features of echo. Then, we reason that an individual dimension or channel in this well-learned latent space will be responsible for generating echo-specific noise. Next, we identify and manipulate the aforementioned noise channel to construct a synthetic dataset of paired clean and noisy images with the objective of training a downstream model for image denoising. Lastly, we conduct both internal and external experiments to demonstrate our model's state-of-the-art performance in comparison to an existing baseline method.

\section{Related Works}
\label{sec:relatedworks}

The application of deep learning for echo denoising is an understudied field mainly due to the infeasibility of collecting noisy and clean image pairs, the difficulty of developing a statistical model to explain echo-specific noise, and the intrinsic noisiness of echo acquisitions. Nonetheless, a handful of works explore the use of deep learning to present potential solutions for denoising echo.

In \cite{diffusion}, Stevens \etal train separate score-based diffusion models \cite{score} to model cardiac tissue and noise in the RF domain and implement joint posterior sampling using guidance terms to separate cardiac tissue from noise. Despite a promising outcome, RF signals are non-interpretable, intermediate representations of raw data inside ultrasound machines that require further processing to produce the familiar B-mode images that visually display cardiac structures and motion. RF signals are invariably discarded in clinical practice due to their lack of clinical relevance. Therefore, the application of this approach to existing acquisitions is fundamentally limited. 

Furthermore, Jahren \etal train a supervised convolutional neural network by constructing a paired dataset of clean and noisy images in an RF-based domain \cite{suppress}. The authors manually excise rectangular patches of noisy regions in echo images and superimpose them onto clean images to acquire noisy image counterparts. Unfortunately, this method shares the same limitation as the approach proposed by Stevens \etal. Furthermore, this excising approach is only feasible for RF-based data which come in rectangular formats. This is in contrast to B-mode images which undergo a scan conversion process to produce a sectoral format.

\section{Methods}
\label{section:methods}

The goal of our method is (1) to develop a generative model capable of simulating ultrasound noise and (2) use this generative model to construct a synthetic dataset of paired clean and noisy images to train a downstream model for image denoising.

\begin{figure*}
\centering
\includegraphics[width=0.99\linewidth]{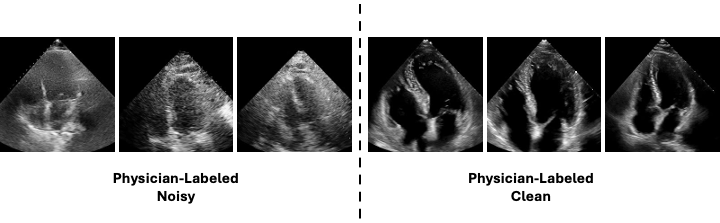}
\caption{Examples of noisy and clean images labeled by expert physicians. Criteria for image quality largely determined by visibility of cardiac structures and absence of noise in chambers.}
\label{figure:physician_examples}
\end{figure*}

\begin{figure*}
\centering
\includegraphics[width=0.99\linewidth]{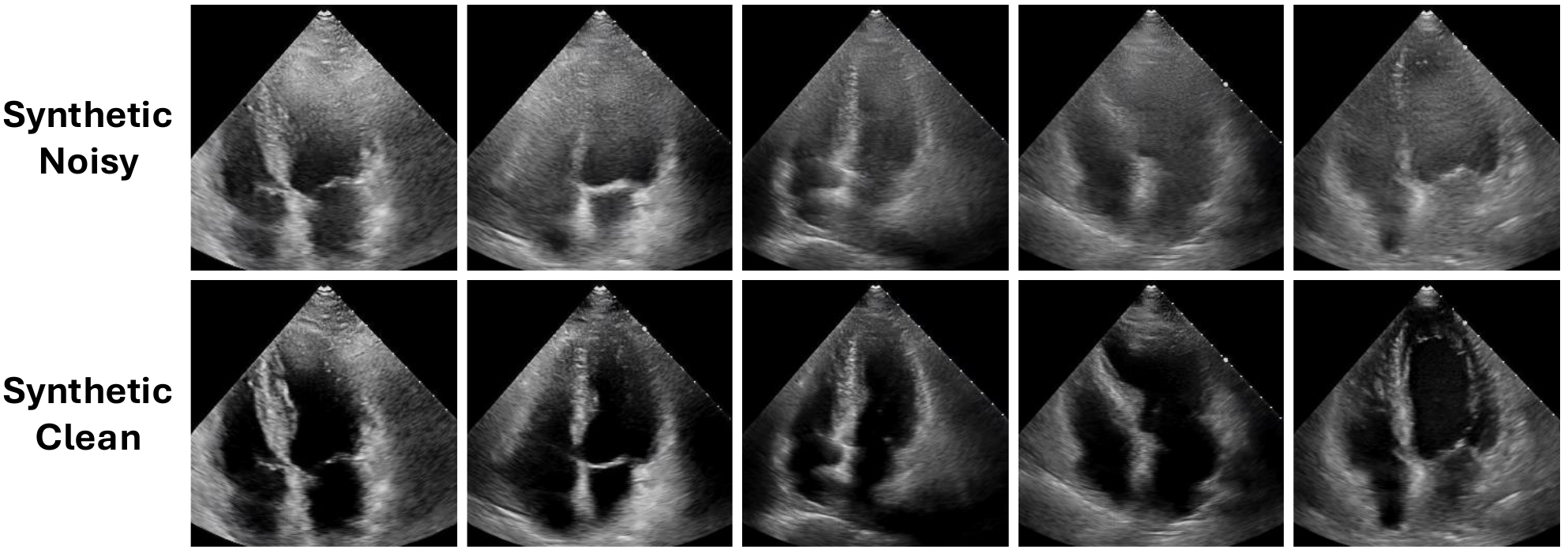}
\caption{Examples of paired noisy and clean images in our synthetic dataset constructed via noise channel manipulation. Note the preservation of depth indicators, sector geometry, and cardiac structures, alongside the restriction of noise to the interior of the sector region.}
\label{figure:synthetic_examples}
\end{figure*}

\subsection{Noise Simulation}
\label{methods:noise_simulation}

In this work, we opt to train a StyleGAN \cite{StyleGAN} to learn a latent space representation $S \subseteq R^{9088}$ \cite{stylespace} for the space of echo images $I \subseteq R^{256\times256}$ and a corresponding synthesis network $G_{s} \colon S \longmapsto I$. In view of the fact that StyleGAN does not have a built-in inversion method, we additionally train an encoder $\mathcal{E} \colon I \longmapsto S$ \cite{encoder} to learn the inverse mapping of echo images to latent embeddings.

We reason that a model which learns to generate echo images must, by natural consequence, learn to generate intrinsic features of echo e.g.\ noise. Furthermore, we infer that individual dimensions or channels of $S$ correspond to independent features that can be manipulated to control specific features in echo images without altering other pre-existing features.

\cref{figure:overview} provides an overview of our noise simulation method. Our key insight is that channels with the highest degree of separation between the clean and noisy distribution are most likely to be responsible for controlling noise. To measure separation, we compute the magnitude of differences between noisy latent embeddings and mean of clean latent embeddings.

To be more specific, we give expert physicians a collection of echo images and ask them to label a subset of clean images $I_c$ and noisy images $I_n$ with examples shown in \cref{figure:physician_examples}. Then, we project $I_c$ and $I_n$ to latent embeddings $S_c = \mathcal{E}(I_c)$ and $S_n = \mathcal{E}(I_n)$. Next, we let $\mu_c$ and $\sigma_c$ denote the mean and standard deviation of $S_c$ and compute the differences $\delta_n = \frac{(S_n - \mu_c)}{\sigma_c}$. Finally, we let $\mu_n$ and $\sigma_n$ denote the mean and standard deviation of $\delta_n$ and compute the magnitude of $\delta_n$ with $\theta_n = \frac{\lvert \mu_n \rvert}{\sigma_n}$.

\begin{figure*}[!ht]
\centering
\includegraphics[width=0.99\linewidth]{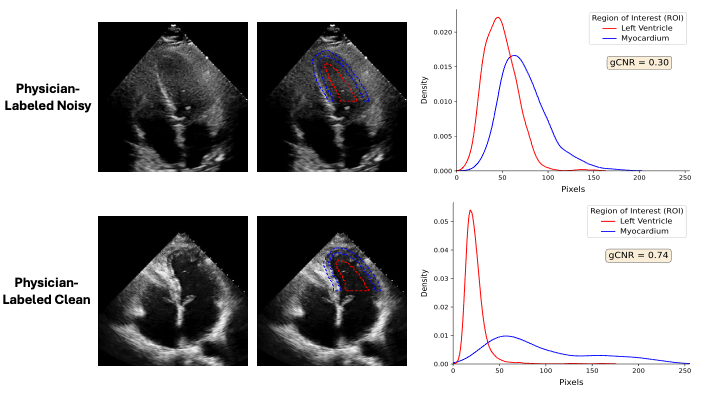}
\caption{Demonstration of gCNR metric. Top and bottom row corresponds to a noisy and clean example, respectively. From left to right, subsequent columns represent a physician-labeled image, regions of interest, and metric computation. Note, an increase in overlap between densities results in a lower gCNR and vice versa.}
\label{figure:metric}
\end{figure*}

\begin{figure*}
\centering
\includegraphics[width=0.99\linewidth]{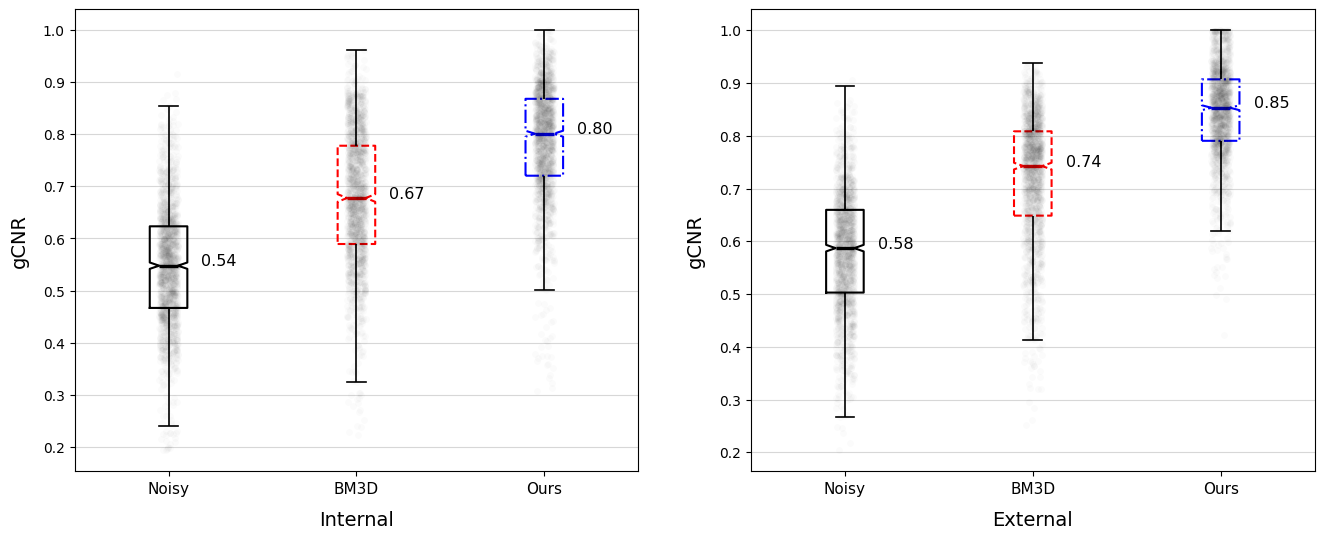}
\caption{Quantitative results for both internal and external test sets. Notched box plots display distribution of gCNR scores for noisy images, baseline BM3D method, and our method. Notches inside boxes represent the 95\% confidence interval about the median.}
\label{figure:comparison_experiment}
\end{figure*}

For each channel in $\theta_n$, the corresponding magnitude indicates the degree of separation between the clean and noisy distribution. To identify the noise channel, we filter the top 100 channels with the highest magnitudes in $\theta_n$ and ask expert physicians to analyze the effect of individually manipulating these channels in some clean latent embedding $s_c$. We define the manipulation of a channel $c_i$ to be $\Delta_i \in S$ where all channels are set to 0 except for $c_i$ and the manipulation effect to be the visual difference between $G_s(s_c)$ and $G_s(s_c + \Delta_i)$. Then, we define $c_{noise}$ and $\Delta_{c \rightarrow n}$ to be the channel and corresponding manipulation that the expert physicians determine to be independently responsible for generating noise.

\begin{figure*}
\centering
\includegraphics[width=0.99\linewidth]{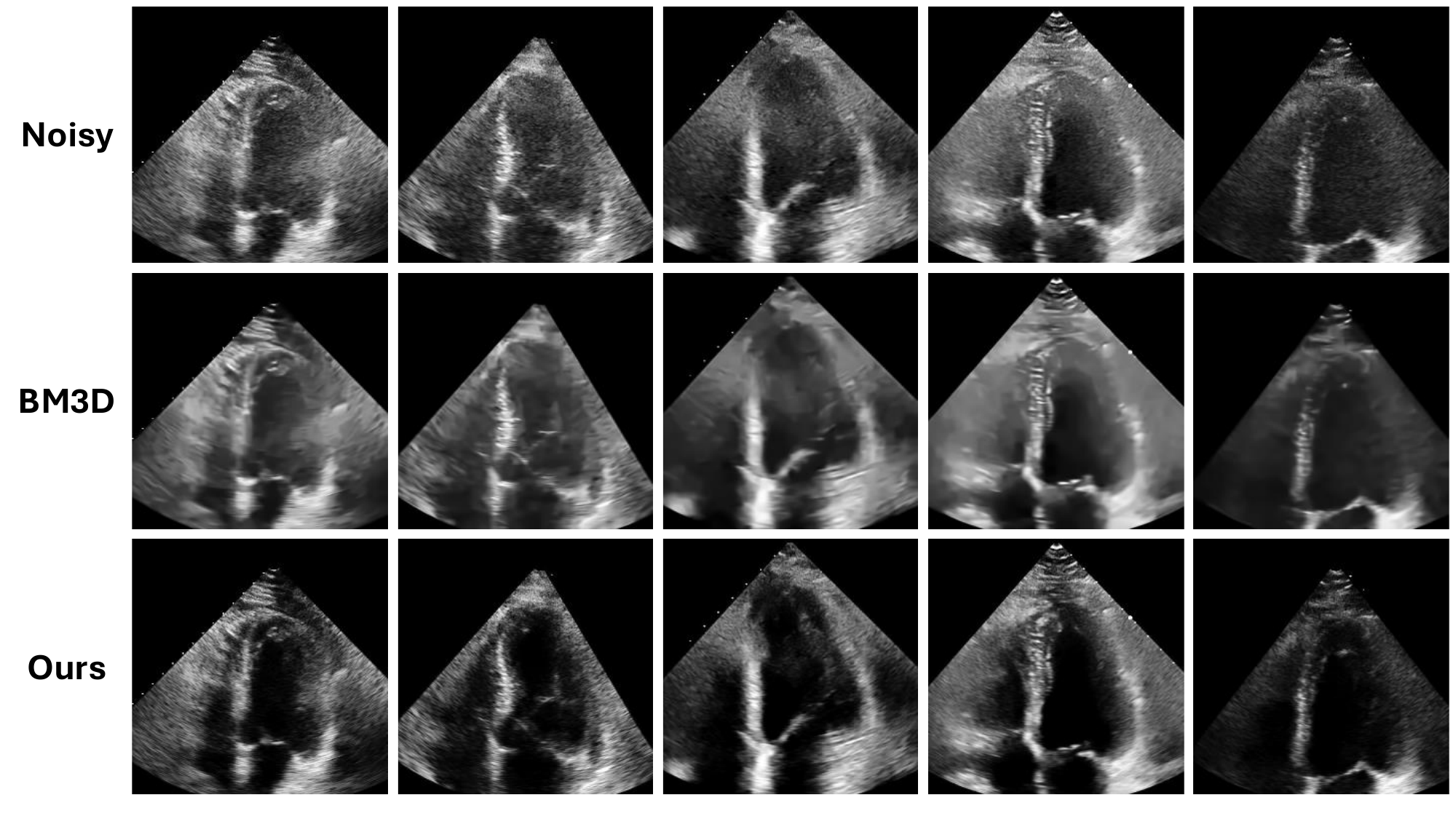}
\caption{Qualitative results for internal set. Top row shows noisy images, middle row shows results from baseline BM3D method, and bottom row shows results from our method.}
\label{figure:internal}
\end{figure*}

\begin{figure*}
\centering
\includegraphics[width=0.99\linewidth]{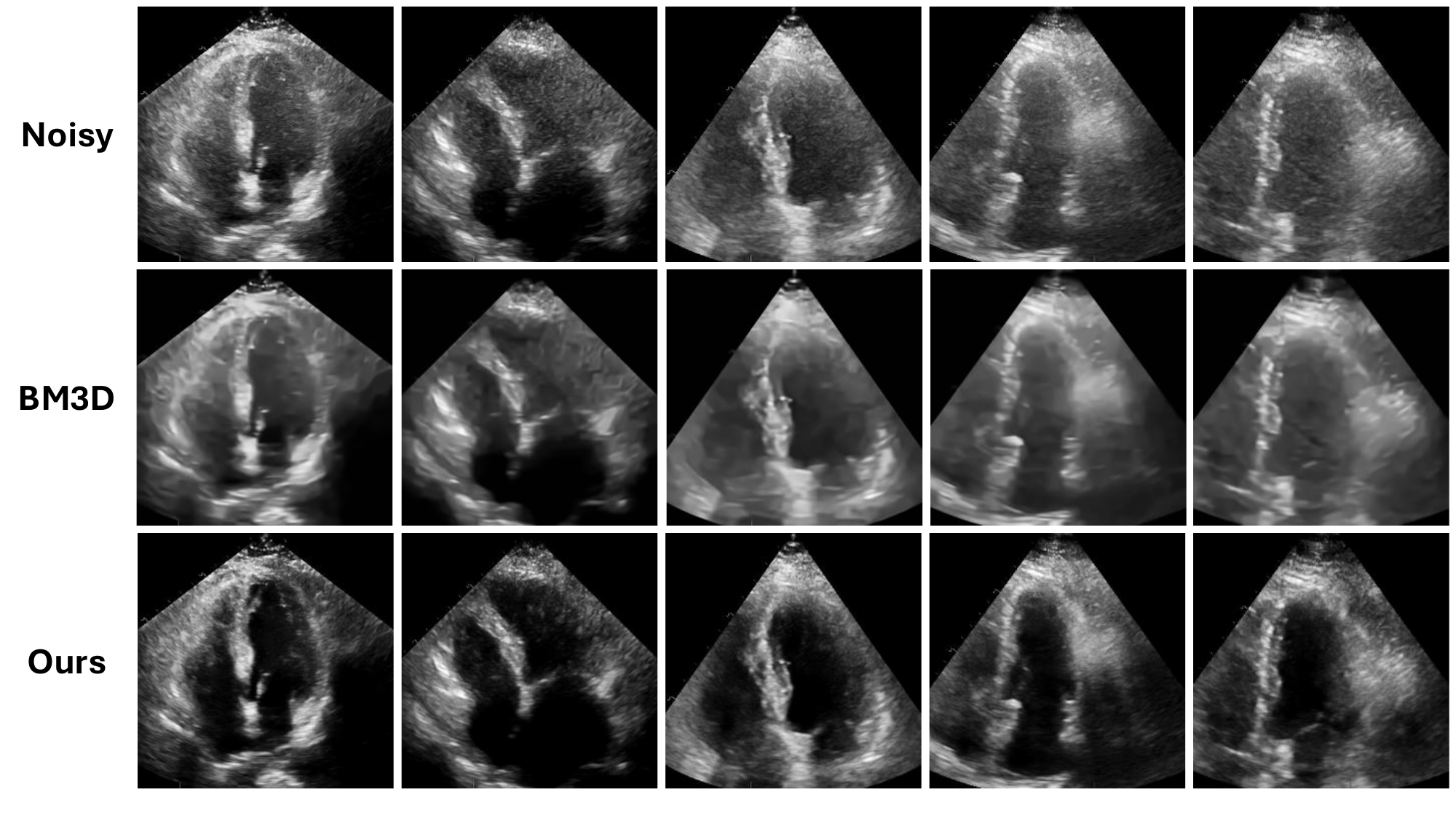}
\caption{Qualitative results for external set. Top row shows noisy images, middle row shows results from baseline BM3D method, and bottom row shows results from our method.}
\label{figure:external}
\end{figure*}

\subsection{Synthetic Dataset}
\label{methods:synthetic_dataset}

To train a downstream model for image denoising, we construct a synthetic dataset of paired noisy and clean images. First, we give expert physicians a collection of echo images and ask them to label a subset of 5600 clean images $I_c$ which we project to latent embeddings $S_c = \mathcal{E}(I_c)$. Then, we use the reconstructions $\widehat{I_c} = G_s(S_c)$ and manipulations $\widehat{I_n} = G_s(S_c + \Delta_{c \rightarrow n})$ as clean and noisy images for our synthetic dataset.

The motivation behind using $\widehat{I_c}$ instead of $I_c$ comes from the fact that GANs utilize an adversarial loss for training as opposed to a reconstruction-based loss. Thus, reconstructed images derived from latent embeddings are not accurate reflections of the corresponding true images. As such, manipulation of latent embeddings results in transformations of reconstructed images as opposed to true images.

 Finally, we implement a standard U-Net \cite{unet} with four residual blocks and channel sizes 64, 128, 256, 512 for our downstream denoising model and train the model with $\widehat{I_n}$ as inputs and $\widehat{I_c}$ as target outputs. \cref{figure:synthetic_examples} shows examples from our synthetic dataset and the capabilities of our noise manipulation method.

\section{Experiments}
\label{section:experiments}

\subsection{Datasets}

For model training, we collect 84,953 apical 4-chamber view (A4C) echo videos from 37,307 patients in Cedars-Sinai Medical Center (CSMC). We extract 5,293,146 images from videos and use a 90-5-5 split for training, validation, and test sets, yielding 4,763,832, 264,657, 264,657 images, respectively. For the downstream denoising task, we ask expert physicians to identify 1,626 noisy images from the CSMC test set for internal evaluation and 1,651 noisy images from the publicly available MIMIC-IV-ECHO \cite{physionet, mimic-iv, mimic-iv-echo, mimic-iv-echo-parent} dataset for external evaluation. For pre-processing, we convert the colorscheme and resolution of all images to grayscale and $256\times256$, respectively.

\subsection{Metric}

In practical real-world denoising, clean references do not exist for noisy images. Thus, it is impossible to rely on the standard image quality metrics, e.g. PSNR, SSIM, MSE, used in controlled settings where a basic noise model is assumed e.g. Gaussian, Poisson, etc. Therefore, we utilize an unsupervised metric known as the generalized contrast-to-noise ratio (gCNR) \cite{gCNR} widely used in ultrasound imaging to measure quality, given by:\begin{equation} \label{equation:gCNR}gCNR(I) = 1 - \int_{x \in I} min \{ p_A(x), p_B(x) \} dx,\end{equation} where $p_A$ and $p_B$ are pixel distributions corresponding to different regions of interest (ROIs) in an image. In practice, the visual distinction between the left ventricle (LV) and myocardium is important for making accurate clinical measurements e.g. LV tracing, LV ejection fraction, etc. Therefore, we decide on using the LV and myocardium as the ROIs.

We utilize the LV segmentation model EchoNet-Dynamic \cite{dynamic} to acquire the LV region. Then, we expand the LV region and remove the original section to acquire the myocardium region. To mitigate the papillary muscles, trabeculations, and mitral valves from biasing our metric, we compress the LV region and truncate 30\% off the bottom. Afterwards, an expert physician reviews the resulting ROIs and removes images with inaccurate results. \cref{figure:metric} shows a demonstration of our metric.

\subsection{Results}

For a baseline comparison of our method's denoising performance, we implement the block-matching and 3D filtering (BM3D) \cite{bm3d} denoising algorithm. BM3D is a standard baseline for evaluating the performance of image denoising methods owing to its flexibility for general application. To the best of our knowledge, there are no existing publicly accessible implementations of deep learning models for the purpose of denoising echo in the B-mode domain.

We provide quantitative comparisons in \cref{figure:comparison_experiment} for both internal and external test sets. In our internal evaluation, we demonstrate that our method achieves a higher median gCNR of $0.80~(95\% ~CI ~0.79 - 0.81)$ compared to BM3D at $0.67~(95\% ~CI ~0.67 - 0.69)$ and original noisy images at $0.54~(95\% ~CI ~0.54 - 0.56)$. In our external evaluation, we likewise report that our method achieves a higher median gCNR of $0.85~(95\% ~CI ~0.85 - 0.86)$ compared to BM3D at $0.74~(95\% ~CI ~0.74 - 0.75)$ and original noisy images at $0.58~(95\% ~CI ~0.58 - 0.59)$.

Furthermore, we provide qualitative comparisons in \cref{figure:internal} and \cref{figure:external} and demonstrate their consistency with our reported quantitative findings. In both internal and external qualitative comparisons, a close examination of denoised images from BM3D shows that the algorithm compromises spatial resolution and fine textural details to smooth out high-frequency noise. Evidently, BM3D is incapable of removing noise in the atrial and ventricular chambers of depicted hearts in B-mode echo. By contrast, our method demonstrates the capability of removing noise in the cardiac chambers without compromising spatial resolution or fine textural detail. We highlight the notable difference in visual distinction between cardiac chambers and structures in the denoised images derived from our method in comparison to those resulting from BM3D.

\section{Conclusions}
\label{section:conclusions}

In this work, we introduce the first deep generative model capable of simulating ultrasound noise and use this generative model to construct a synthetic dataset of paired clean and noisy images for training a downstream model for image denoising. We believe our work to be a significant step forward in the field of medical imaging for denoising tasks, proving that independent modeling of modality-specific noise is without a doubt possible. Currently, the limitation of our work comes from the fact that our method is image-based and does not exploit temporal information. With computer vision research shifting towards video-based generative models, a potential direction for future research involves extending our work to videos using diffusion-based architectures and techniques. Nonetheless, we trust that our work will contribute to a more equitable healthcare and research environment with implications of increasing data availability and improving diagnostics without relying on advanced equipment.

{
    \small
    \bibliographystyle{ieeenat_fullname}
    \bibliography{main}
}

\end{document}